\documentclass[11pt,letterpaper]{article}
\usepackage{amsmath}
\usepackage{amssymb}
\usepackage{fullpage}
\usepackage[amsthm,amsmath,thmmarks]{ntheorem}

\newtheorem{theorem}{Theorem}
\newtheorem{corollary}{Corollary}
\newtheorem{lemma}{Lemma}
\newtheorem{claim}{Claim}
\newtheorem*{definition}{Definition}

\newcommand{\TC}{{\rm TC}}

\newcommand{\BPTISP}{{\rm BPTISP}}
\newcommand{\BQTISP}{{\rm BQTISP}}
\newcommand{\PTISP}{{\rm PTISP}}
\newcommand{\BQTIME}{{\rm BQTIME}}
\newcommand{\PTIME}{{\rm PTIME}}
\newcommand{\p}{{\rm P}}
\newcommand{\NP}{{\rm NP}}
\newcommand{\PH}{{\rm PH}}
\newcommand{\PARITYP}{{\rm \oplus P}}
\newcommand{\BQP}{{\rm BQP}}
\newcommand{\PP}{{\rm PP}}
\newcommand{\SHARPP}{{\rm \#P}}

\newcommand{\SAT}{{\rm SAT}}
\newcommand{\MAJSAT}{{\rm MajSAT}}
\newcommand{\MAJMAJSAT}{{\rm MajMajSAT}}

\newcommand{\poly}{{\rm poly}}
\newcommand{\polylog}{{\rm polylog}}

\newcounter{backup1}
\newcounter{backup2}

\begin{document}

\title{A Quantum Time-Space Lower Bound for the Counting 
Hierarchy\thanks{Research partially supported by NSF awards
CCR-0133693 and CCF-0523680.}}
\author{Dieter van Melkebeek \hspace{0.5in} Thomas Watson \vspace{0.05in}\\
University of Wisconsin - Madison}
\maketitle

\begin{abstract}
We obtain the first nontrivial time-space lower bound for quantum algorithms
solving problems related to satisfiability. Our bound applies to $\MAJSAT$ and
$\MAJMAJSAT$, which are complete problems for the first and second levels of
the counting hierarchy, respectively. We prove that for every real $d$ and
every positive real $\epsilon$ there exists a real $c>1$ such that either:
\begin{itemize}
\item $\MAJMAJSAT$ does not have a quantum algorithm with bounded two-sided
error that runs in time $n^c$, or
\item $\MAJSAT$ does not have a quantum algorithm with bounded two-sided error
that runs in time $n^d$ and space $n^{1-\epsilon}$.
\end{itemize}
In particular, $\MAJMAJSAT$ cannot be solved by a quantum algorithm with
bounded two-sided error running in time $n^{1+o(1)}$ and space $n^{1-\epsilon}$
for any $\epsilon>0$.

The key technical novelty is a time- and space-efficient simulation of quantum
computations with intermediate measurements by probabilistic machines with
unbounded error. We also develop a model that is particularly suitable for the
study of general quantum computations with simultaneous time and space bounds.
However, our arguments hold for any reasonable uniform model of quantum
computation.
\end{abstract}


\section{Introduction}
\label{qts:sec:intro}

Satisfiability, the problem of deciding whether a given Boolean formula has
at least one satisfying assignment, has tremendous practical and theoretical
importance. It emerged as a central problem in complexity theory with the
advent of $\NP$-completeness in the 1970's. Proving lower bounds on the
complexity of satisfiability remains a major open problem. Complexity theorists
conjecture that satisfiability requires exponential time and linear space to
solve in the worst case. Despite decades of effort, the best single-resource
lower bounds for satisfiability on general-purpose models of computation are
still the trivial ones -- linear for time and logarithmic for space. However,
since the late 1990's we have seen a number of results that rule out certain
nontrivial combinations of time and space complexity.

One line of research \cite{For, FLvMV, Wil1, DvM, Wil2}, initiated by Fortnow,
focuses on proving stronger and stronger time lower bounds for deterministic
algorithms that solve satisfiability in small space. For subpolynomial (i.e.,
$n^{o(1)}$) space bounds, the current record states that no such algorithm can
run in time $n^c$ for any $c<2\cos{(\pi/7)}\approx 1.8019$. A second research
direction aims to strengthen the lower bounds by considering more powerful
models of computation than the standard deterministic one. Diehl and Van
Melkebeek \cite{DvM} initiated the study of lower bounds for problems related
to satisfiability on randomized models with bounded error. They showed that
for every integer $\ell\geq 2$, $\Sigma_\ell\SAT$ cannot be solved in time
$n^c$ by subpolynomial-space randomized algorithms with bounded two-sided
error for any $c<\ell$, where $\Sigma_\ell\SAT$ denotes the problem of
deciding the validity of a given fully quantified Boolean formula with $\ell$
alternating blocks of quantifiers beginning with an existential quantifier.
$\Sigma_\ell\SAT$ represents the analogue of satisfiability for the $\ell$th
level of the polynomial-time hierarchy; $\Sigma_1\SAT$ corresponds to
satisfiability. Proving nontrivial time-space lower bounds for satisfiability
on randomized algorithms with bounded two-sided error remains open. Allender
et al.~\cite{AKRRV} considered the even more powerful (but physically
unrealistic) model of probabilistic algorithms with unbounded
error\footnote{Throughout this paper, we use the word ``randomized'' for the
bounded-error setting and ``probabilistic'' for the unbounded-error setting.}.
They settled for problems that are even harder than $\Sigma_\ell\SAT$ for any
fixed $\ell$, namely $\MAJSAT$ and $\MAJMAJSAT$, the equivalents of
satisfiability and $\Sigma_2\SAT$ in the counting hierarchy. $\MAJSAT$ is the
problem of deciding whether a given Boolean formula is satisfied for at least
half of the assignments to its variables. $\MAJMAJSAT$ is the problem of
deciding whether a given Boolean formula $\varphi$ on disjoint variable sets
$x$ and $y$ has the property that for at least half of the assignments to $x$,
$\varphi$ is satisfied for at least half of the assignments to $y$. Recall
that Toda \cite{Toda} proved that the polynomial-time hierarchy reduces to the
class $\PP$, which represents polynomial-time probabilistic computations with
unbounded two-sided error and forms the first level of the counting hierarchy.
Apart from dealing with harder problems, the quantitative strength of the
lower bounds by Allender et al.\ is also somewhat weaker. In particular, they
showed that no probabilistic algorithm can solve $\MAJMAJSAT$ in time
$n^{1+o(1)}$ and space $n^{1-\epsilon}$ for any positive constant $\epsilon$.
We refer to \cite{vM} for a detailed survey of the past work on time-space
lower bounds for satisfiability and related problems, including a presentation
of the Allender et al.\ lower bound that is slightly different from the
original one.

In this paper we study the most powerful model that is considered physically
realistic, namely quantum algorithms with bounded error. We obtain the first
nontrivial time-space lower bound for quantum algorithms solving problems
related to satisfiability. In the bounded two-sided error randomized setting,
the reason we can get lower bounds for $\Sigma_\ell\SAT$ for $\ell\geq 2$ but
not for $\ell=1$ relates to the fact that we know efficient simulations of
such randomized computations in the second level of the polynomial-time
hierarchy but not in the first level. In the quantum setting the situation is
worse: we know of no efficient simulations in any level of the polynomial-time
hierarchy. The best simulations to date are due to Adleman et al.~\cite{ADH},
who showed that polynomial-time quantum computations with bounded two-sided
error can be simulated in $\PP$. Building on this connection, we bring the
lower bounds of Allender et al.\ to bear on bounded-error quantum algorithms.
Our main result shows that either a time lower bound holds for quantum
algorithms solving $\MAJMAJSAT$ or a time-space lower bound holds for
$\MAJSAT$.
\begin{theorem}
\label{qts:thm:maintheorem}
For every real $d$ and every positive real $\epsilon$ there exists a real
$c>1$ such that either:
\begin{itemize}
\item $\MAJMAJSAT$ does not have a quantum algorithm with bounded two-sided
error that runs in time $n^c$, or
\item $\MAJSAT$ does not have a quantum algorithm with bounded two-sided error
that runs in time $n^d$ and space $n^{1-\epsilon}$.
\end{itemize}
\end{theorem}
As a corollary, we obtain a single time-space lower bound for $\MAJMAJSAT$.
\begin{corollary}
\label{qts:cor:maincorollary}
$\MAJMAJSAT$ cannot be solved by a quantum algorithm with bounded two-sided
error running in time $n^{1+o(1)}$ and space $n^{1-\epsilon}$ for any
$\epsilon>0$.
\end{corollary}

Unlike in the deterministic and randomized cases, it is not obvious how to
define a model of quantum computation that allows us to accurately measure
both time and space complexity. The existing models give rise to various 
issues. For example, intermediate measurements play a critical role as
they are needed for time-space efficient simulations of randomized
computations by quantum computations. Several of the known models only allow
measurements at the end of the computation but not during the computation. As
another example, the classical time-space lower bounds hold for models with
random access to the input and memory. This makes the lower bounds more
meaningful as they do not exploit artifacts due to sequential access.
Extending the standard quantum Turing machine model \cite{BV} to accommodate
random access leads to complications that make the model inconvenient to work
with. In Section \ref{qts:sec:model} we discuss these and other issues in
detail, and we survey the known models from the literature. We present a model
that addresses all issues and is capable of efficiently simulating all other
uniform models that may be physically realizable in the foreseeable future.
Thus, lower bounds in our model reflect true problem hardness.

The main technical novelty for establishing Theorem \ref{qts:thm:maintheorem}
consists of an improved time- and space-efficient simulation of quantum
computations by unbounded-error probabilistic computations. The previously
known simulations, such as the one by Adleman et al., do not deal with
intermediate measurements in a space-efficient way. We show how to cope with
intermediate measurements in a space-efficient way and without loss in running
time. Our construction works even when the sequence of local quantum
operations can depend on previous measurement outcomes; i.e., we handle more
powerful models than uniform quantum circuits. Our simulation makes use of a
result on approximating quantum gates due to Solovay and Kitaev \cite{Kit}.
Theorem \ref{qts:thm:maintheorem} follows from our simulation and the Allender
et al.\ lower bound. The quantitative strength of our lower bound derives from
the latter; our translation does not induce any further weakening.

The rest of this paper is organized as follows. We start with a discussion
of the model in Section \ref{qts:sec:model}, we derive our results in
Section \ref{qts:sec:lowerbound}, and we conclude with some open problems
in Section \ref{qts:sec:concl}. Throughout we assume basic background in
quantum computation; see for example \cite{NC, KSV}.


\section{Models of Quantum Computation}
\label{qts:sec:model}

In this section we develop the model that we use for the exposition of our
arguments. Section \ref{qts:sec:modelissues} contains a discussion
of the issues that arise in choosing a model of quantum computation that
accurately reflects time and space complexity. In Section
\ref{qts:sec:earliermodels} we describe how previously studied models fit into
our taxonomy. We motivate and precisely define our chosen model in Section
\ref{qts:sec:modeldefinition}. Although we consider the development of such a
model as a contribution of our paper, the crux of our main result can be
understood at an abstract level. As such, a reader who would like to quickly
get to the heart of our paper can skip Section 2.


\subsection{Issues}
\label{qts:sec:modelissues}

Our model should capture the notion of a quantum algorithm as viewed by the
computer science and physics communities and allow us to accurately measure the
resources of time and space. For example, the model should allow us to express
important quantum algorithms such as Shor's \cite{Shor} and Grover's
\cite{Grov} in a way that is natural and faithfully represents their
complexities. This forms the overarching issue in choosing a model. Below we
discuss eight specific aspects of quantum computation models and describe how
the corresponding issues are handled in the classical setting.\bigskip

\noindent {\bf Sublinear space bounds.} Many algorithms have the property that
the amount of work space needed is less than the size of the input. Models
such as one-tape Turing machines do not allow us to accurately measure the
space usage of such algorithms because they charge for the space required to
store the input. In the deterministic and randomized settings, sublinear space
bounds are accommodated by considering Turing machines with a read-only input
tape that does not count toward the space bound and read-write work tapes that
do. In the quantum setting, we need a model with an analogous
capability.\bigskip

\noindent {\bf Random access to the input and memory.} In order to accurately
reflect the complexity of computational problems, our model should include a
mechanism for random access, i.e., the ability to access any part of the input
or memory in a negligible amount of time (say, linear in the length of the
address). For example, there is a trivial algorithm for the language of
palindromes that runs in quasilinear time and logarithmic space on standard
models with random access, but the time-space product of any traditional
sequential-access Turing machine deciding palindromes is at least quadratic.
The latter result does not reflect the complexity of deciding palindromes, but
rather exploits the fact that sequential-access machines may have to waste a
lot of time moving their tape heads back and forth. Classical Turing machines
can be augmented with a mechanism to support random access; our quantum model
should also have such a mechanism.\bigskip

\noindent {\bf Intermediate measurements.} Unlike the previous two issues,
intermediate measurements are specific to the quantum setting. In time-bounded
quantum computations, it is customary to assume that all measurements occur at
the end. This is because intermediate measurements can be postponed by
introducing ancilla qubits to store what would be the result of the
measurement, thus preventing computation paths with different measurement
outcomes from interfering with each other. However, this has a high cost in
space -- a computation running in time $t$ may make up to $t$ measurements, so
the space overhead could be as large as $t$, which could be exponential in the
original space bound. Hence, to handle small space bounds our model should
allow intermediate measurements. Indeed, this is crucial for our model to meet
the expectation of being at least as strong as randomized algorithms with
comparable efficiency parameters; the standard way to ``flip a coin'' in the
quantum setting is to apply a Hadamard gate to a qubit in a basis state and
then measure it. Also, many quantum algorithms, such as Shor's factoring
algorithm, are naturally described using intermediate measurements.

We also need to decide which measurements to allow. Projective measurements in
the computational basis are the most natural choice. Should we allow projective
measurements in other bases? How about fully general measurements (see Section
2.2.3 in \cite{NC}), where the measurement operators need not be projections?
General measurements can be performed by introducing ancilla qubits (at a cost
in space), performing a change of basis (at a cost in time), and doing a
projective measurement in the computational basis, one qubit at a time. It is
reasonable to charge the complexity of these operations to the algorithm
designer, so we are satisfied with allowing only single-qubit measurements in
the computational basis.\bigskip

\noindent {\bf Obliviousness to the computation history.} Computations proceed
by applying a sequence of local operations to data. We call a computation
\emph{nonoblivious} if at each step, which local operation to use and which
operands to apply it to may depend on the computation history. A generic
deterministic Turing machine computation is nonoblivious. We can view each
state as defining an operation on a fixed number of tape cells, where the
operands are given by the tape head locations. In each step, the outcome of the
applied operation affects the next state and tape head locations, so both the
operation and the operands can depend on the computation history. In contrast,
a classical circuit computation is oblivious because neither the operation
(gate) nor the operands (wires connected to the gate inputs) depend on the
computation history (values carried on the wires).

In the randomized and quantum settings, the notion of a computation history
becomes more complicated because there can be many computation paths. In the
randomized setting, applying a randomized operation to a configuration may
split it into a distribution over configurations, and the randomized Turing
machine model allows the next state and tape head locations to depend on which
computation path was taken. In the quantum setting, applying a quantum
operation to a basis state may split it into a superposition over several basis
states, and general nonoblivious behavior would allow the next operation and
operands to depend on which computation path was taken. However, it is unclear
whether such behavior is physically realizable, as currently envisioned
technologies all select quantum operations classically. An intermediate notion
of nonobliviousness, where the operations and operands may depend on previous
measurement outcomes but not on the quantum computation path, does seem
physically realistic.\bigskip

\noindent {\bf Classical control.} There is a wide spectrum of degrees of
interaction between a quantum computation and its classical control. On the
one hand, one can imagine a quantum computation that is entirely
``self-sufficient,'' other than the interaction needed to provide the input
and observe the output. On the other hand, one can imagine a quantum
computation that is guided classically every step of the way. Self-sufficiency
is inherent to computations that are nonoblivious to the quantum computation
path, whereas measurements are inherently classically controlled operations.
Incorporating intermediate measurements into computations that are
nonoblivious to the quantum computation path would require some sort of global
coordination among the quantum computation paths to determine when a
measurement should take place.\bigskip

\noindent {\bf Syntax.} Our model should be syntactic, meaning that identifying
valid programs in the model is decidable. If we are interested in bounded-error
computations, then we cannot hope to decidably distinguish programs satisfying
the bounded-error promise from those that do not. However, we should be able to
distinguish programs that evolve according to the postulates of quantum
mechanics from those that do not. Allowing nonobliviousness to the quantum
computation path complicates this syntax check. If different components of the
superposition can undergo different unitary operations then the overall
operation is not automatically unitary, due to interference. Extra conditions
on the transition function are needed to guarantee unitarity.\bigskip

\noindent {\bf Complexity of the transition amplitudes.} Care should be taken
in specifying the allowable transition amplitudes. In the randomized setting,
it is possible to solve undecidable languages by encoding the characteristic
sequences of these languages in the transition probabilities. This problem is
usually handled by using a certain universal set of elementary randomized
operations, e.g., an unbiased coin flip. In the quantum setting, the same
problem arises with unrestricted amplitudes. Again, one can solve the problem
by restricting the elementary quantum operations to a universal set. However,
unlike in the randomized setting, there is no single standard universal set
like the unbiased coin flip with which all quantum algorithms are easy to
describe. Algorithm designers should be allowed to use arbitrary local
operations provided they do not smuggle hard-to-compute information into the
amplitudes.\bigskip

\noindent {\bf Absolute halting.} In order to measure time complexity, we
should use a model that naturally allows any algorithm to halt absolutely
within some time bound $t$. In the randomized setting, one can design
algorithms whose running times are random variables and may actually run
forever. We can handle such algorithms by clocking them, so that they are
forced to halt within some fixed number of time steps. Our quantum model should
provide a similar mechanism.\bigskip


\subsection{Earlier Models}
\label{qts:sec:earliermodels}

Now that we have spelled out the relevant issues and criteria, we consider
several previously studied models as candidates.

Bernstein and Vazirani \cite{BV} laid the foundations for studying quantum
complexity theory using quantum Turing machines. Their model uses a single
tape and therefore cannot handle sublinear space bounds. Like classical
one-tape Turing machines, their model is sequential-access. It does not allow
intermediate measurements. On the other hand, their model is fully
nonoblivious: the transition function produces a superposition over basis
configurations, and the state and tape head location may be different for
different components of the superposition. Their model represents the
self-sufficient extreme of the classical control spectrum. In their paper,
Bernstein and Vazirani prove that their model is syntactic by giving a
few orthogonality constraints on the entries of the transition function table
that are necessary and sufficient for the overall evolution to be unitary.
These conditions are somewhat unnatural, and can be traced back to the
possibility of nonobliviousness to the quantum computation path. Bernstein and
Vazirani restrict the transition amplitudes by requiring that the first $k$
bits of each amplitude are computable deterministically in time $\poly(k)$.
Their model is nontrivial to clock; they require that the transition function
be designed in such a way that the machine always halts, meaning that it
reaches a superposition in which all non-halting basis configurations have
zero amplitude. Bernstein and Vazirani detail how to design such mechanisms.

In \cite{Wat1}, Watrous considers a model similar to Bernstein and Vazirani's,
but with one read-write work tape and a read-only input tape not counting
toward the space bound. The model naturally allows for sublinear space bounds,
but it is still sequential-access. It allows intermediate measurements but
only for the halting mechanism: a special register is measured after each time
step, with the outcome indicating ``halt and output 1'', ``halt and output
0'', or ``continue''. The model is nonoblivious like the Bernstein-Vazirani
model. It has more classical interaction due to the halting mechanism, but
this is arguably not ``classical control.'' The syntax conditions on the
transition function are similar to those for the Bernstein-Vazirani model. The
results in \cite{Wat1} require the transition amplitudes to be rational, which
is somewhat unappealing since one may often wish to use Hadamard gates, which
have irrational amplitudes. Similar to the Bernstein-Vazirani model, the model
is nontrivial to clock. In fact, the results in \cite{Wat1} rely on counting
an infinite computation as a rejection.

The main issue with the above models for our purposes is their
sequential-access nature. It is possible to handle this problem by imposing a
random-access mechanism. However, the conditions on the entries of the
transition function table characterizing unitary evolution become more
complicated and unnatural, making the model inconvenient to work with. Again,
the culprit is the nonobliviousness to the quantum computation path.
Since this behavior does not appear to be physically realizable in the
foreseeable future anyway, the complications arising from it are in some sense
unjustified.\bigskip

In \cite{Wat2}, Watrous considers a different model of space-bounded quantum
computation. This model is essentially a classical Turing machine with an
additional quantum work tape and a fixed-size quantum register. Sublinear
space bounds are handled by charging for the space of the classical work tape
and the quantum work tape but not the input tape. All three tape heads move
sequentially. This model handles intermediate measurements. It is oblivious to
the quantum computation path; the state and tape head locations cannot be in
superposition with the contents of the quantum work tape. However, the
computation is nonoblivious to the classical computation history, including
the measurement outcomes. The finite control is classical; in each step it
selects a quantum operation and applies it to the combination of the qubit
under the quantum work tape head together with the fixed-size register. The
register is needed because there is only one head on the quantum work tape,
but a quantum operation needs to act on multiple qubits to create
entanglement. The allowed operations come from the so-called quantum
operations formalism (see Chapter 8 of \cite{NC}), which encompasses unitary
operations and general measurements, as well as interaction with an external
environment. Each quantum operation produces an output from a finite alphabet
-- the measurement outcome in the case of a measurement. This outcome
influences the next (classical) transition. This model is syntactic just like
classical Turing machines, with the additional step of testing that each
quantum operation satisfies the definition of a valid quantum operation.
For his constructions, Watrous needs the transition amplitudes to be
algebraic. This model is trivial to clock, since all the control is done
classically and thus the machine can halt in a fixed number of steps, just as
in the classical setting.

The latter model is convenient to work with since the essence of the quantum
aspects of a computation are isolated into local operations that are chosen
classically and applied to a simple quantum register. This models the
currently envisioned realizations of quantum computers. We adopt this model
for the exposition of our results, but we need to make some modifications in
order to address the following issues.
\begin{itemize}
\item Algorithms like Grover's require quantum access to the input, i.e.,
an operation that allows different basis states in a superposition to access
different bits of the input simultaneously. On inputs of length $n$, this is
done with a query gate that effects the transformation $|i\rangle|b\rangle
\mapsto|i\rangle|b\oplus x_i\rangle$ where $i\in\{0,1\}^{\lceil\log{n}
\rceil}$, $b\in\{0,1\}$, and $x_i$ is the $i$th bit of the input. The model
from \cite{Wat2} does not have such an operation and thus cannot express
algorithms like Grover's. While this operation seems no more physically
realistic than nonobliviousness to the quantum computation path if we view the
input as stored in a classical memory, it does make sense when the input is
actually the output of another computation. For these reasons, we include such
an operation in our model.

\item We want our model to have random access to emphasize the fact that our
time-space lower bound does not exploit any model artifacts due to sequential
access. We can make the model from \cite{Wat2} random-access by allowing each
of the tape heads to jump in unit time to a location whose address we have
classically computed, just as can be done for deterministic and randomized
Turing machines.

\item The quantum operations used in the model from \cite{Wat2} are more
general than we wish to consider. Since we are focusing on the computational
aspects of the model, we choose to restrict the set of allowed operations to
unitary operations and projective measurements in the computational basis.
The quantum operations formalism models the evolution of open quantum systems,
which is of information-theoretic rather than algorithmic concern and can be
simulated with unitary operations by introducing an additional ``environment''
system at a cost in space.

\item The restriction to algebraic transition amplitudes is unnecessary in the
present setting. We feel that a reasonable way to restrict the amplitudes is
the one chosen by Bernstein and Vazirani; i.e., the first $k$ bits of each
amplitude should be computable deterministically in time $\poly(k)$.
\end{itemize}


\subsection{Our Model}
\label{qts:sec:modeldefinition}

For concreteness, we now describe and motivate the particular model we use
for the exposition of our arguments. Our model addresses all the issues listed
in Section \ref{qts:sec:modelissues}, and is an adaptation of Watrous's model
from \cite{Wat2}, as described at the end of Section
\ref{qts:sec:earliermodels}.

In terms of obliviousness, our model corresponds to the physically realistic
middle ground where a classical mechanism determines which quantum operation
to apply based on the previous measurement outcomes but independent of the
actual quantum computation path. Our arguments are robust with respect to the
details of the model as long as it has the latter property. In particular, we
can handle uniform quantum circuits. Our results also hold for more general
models allowing nonobliviousness to the quantum computation path, but this
requires more technical work; see the remarks in Section
\ref{qts:sec:remarks}.


\subsubsection{Model Definition}

We define a quantum Turing machine as follows. There are three semi-infinite
tapes: the input tape, the classical work tape, and the quantum work tape.
Each cell on the input tape holds one bit or a blank symbol. Each cell on the
classical work tape holds one bit. Each cell on the quantum work tape holds one
qubit. The input tape contains the input, a string in $\{0,1\}^n$, followed by
blanks, and the classical and quantum work tapes are initialized to all $0$'s.
There are a fixed number of tape heads, each of which is restricted to one of
the three tapes. There may be multiple heads moving independently on the same
tape.

The finite control, the operations on the classical work tape, and all head
movements are classical; each operation on the quantum work tape can be either
a unitary operation or a single-qubit projective measurement in the
computational basis. In each step of the computation, the finite control of the
machine is in one of a finite number of states. Each state has an associated
classical function, which is applied to the contents of the cells under the
heads on the classical work tape, and an associated quantum operation, which
is applied to the contents of the cells under the heads on the quantum work
tape. The next state of the finite control and the head movements are
determined by the current state, the contents of the cells under the input
tape heads and classical work tape heads at the beginning of the computation
step, and the measurement outcome if the quantum operation was a measurement.

Each head moves left one cell, moves right one cell, stays where it is, or
jumps to a new location at a precomputed address that is written on the
classical work tape between two of the classical work tape heads. The latter
type of move is classical random access. We also allow ``quantum random
access'' to the input by optionally performing a query that effects the
transformation $|i\rangle|b\rangle\mapsto|i\rangle|b\oplus x_i\rangle$ on the
qubits between two of the quantum work tape heads, where $i\in\{0,1\}^*$ is an
address located on the quantum work tape, $b\in\{0,1\}$, and $x_i$ is the
$i$th bit of the input of length $n$ or $0$ if $i>n$.

Among the states of the finite control are an ``accept'' state and a ``reject''
state, which cause the machine to halt. Although not needed in this paper, the
machine can be augmented with a one-way sequential-access write-only classical
output tape in order to compute nonboolean functions.\bigskip

Let us motivate our model definition. In terms of physical computing systems,
the input tape corresponds to an external input source, the classical work
tape corresponds to classical memory, and the quantum work tape corresponds to
quantum memory. The bits and qubits under the heads correspond to the data
being operated on in the CPU.

We use multiple heads on each tape for several reason. One reason is that
creating entanglement requires multiple-qubit operations and hence multiple
quantum work tape heads. Another reason is that having multiple heads offers a
convenient way of formalizing random access. Since we are studying how
algorithm performance scales with the input size, addresses have non-constant
length and thus cannot fit under the tape heads all at once. A number of
mechanisms are possible for indicating where an address is stored for random
access. The one we have chosen, namely that the address is delimited by two
tape heads, is artificial and is chosen only for convenience because it makes
the model simple and clean. Another possible mechanism is to associate with
each head a special \emph{index tape} used for writing addresses; see
\cite{vM} for a discussion of this type of model.

A minor issue arises with our multiple head approach: an operation on a work
tape may not be well-defined if two of the heads are over the same cell.
Rather than requiring programs to avoid this situation, which would make the
model non-syntactic, we can just assume that no operation is performed on the
violating work tape when this situation arises.

We allow the heads to move sequentially because if we only allowed random
access, then constructing an address would require storing a pointer to the
location where that address is stored. The pointer would have have nonconstant
size, so we would need a pointer to that pointer, and so on. This
chicken-and-egg problem does not appear in physical computing systems, and we
explicitly avoid it by allowing sequential traversal of memory without having
to ``remember'' where the head is.


\subsubsection{Complexity Classes}
\label{qts:sec:classes}

The running time of a quantum Turing machine at input length $n$ is the maximum
over all inputs of length $n$ and over all computation paths of the number of
steps before the machine halts. The space usage is the maximum over all inputs
of length $n$ and over all computation paths of the largest address of a
classical work tape head or quantum work tape head during the computation.
Either the time or the space may be infinite. Note that we maximize over
\emph{all} computation paths, even ones that occur with probability $0$ due to
destructive interference.

Our definition of space usage allows the space to be exponential in the running
time, since in time $t$ a machine can write an address that is exponential in
$t$ and move a head to that location using the random-access mechanism.
However, the space usage can be reduced to at most the running time with at
most a polylogarithmic factor increase in the latter by compressing the data
and using an appropriate data structure to store (old address, new address)
pairs. (See Section 2.3.1 of \cite{vM} for a similar construction.)

We are now set up to define quantum complexity classes within our model.

\begin{definition}
$\BQTISP(t,s)$ is the class of languages $L$ such that for some quantum Turing
machine $M$ running in time $O(t)$ and space $O(s)$,
\begin{itemize}
\item if $x\in L$ then $\Pr(\text{$M$ accepts $x$})\geq\frac{2}{3}$, and
\item if $x\not\in L$ then $\Pr(\text{$M$ accepts $x$})\leq\frac{1}{3}$.
\end{itemize}
We also require that for each entry in the matrix representation of each
unitary operation of $M$ in the computational basis, the first $k$ bits of the
real and imaginary parts are computable deterministically in time $\poly(k)$.
We define $\BQTIME(t)$ similarly but without the space restriction.
\end{definition}

As evidence in support of our model of choice, we note that the following
results hold in our model.
\begin{itemize}
\item Recall that $\BPTISP(t,s)$ is the class of languages solvable in time
$O(t)$ and space $O(s)$ by a randomized algorithm with error probability at
most $1/3$. Then $\BPTISP(t,s)\subseteq\BQTISP(t,s)$ holds because a quantum
algorithm in our model can directly simulate a randomized algorithm; the only
issue is producing unbiased coin flips. For this, the simulation can apply a
Hadamard gate to one qubit on the quantum work tape and then measure it. This
qubit can be reused to generate as many random bits as needed.

\item Grover's algorithm shows that $\text{OR}\in\BQTISP(n^{1/2}\cdot
\polylog(n),\ \log{n})$, where OR denotes the problem of computing the
disjunction of the $n$ input bits.

\item Shor's algorithm shows that a nontrivial factor of an integer of bit
length $n$ can be computed in time $O(n^3\cdot\polylog(n))$ and space $O(n)$
with error probability at most $1/3$ in our model.
\end{itemize}


\section{Time-Space Lower Bound}
\label{qts:sec:lowerbound}

In this section we prove our results. Section \ref{qts:sec:result} contains an
outline of the two main steps of the proof. In Section \ref{qts:sec:approx}
we argue that we can restrict our attention to a special case of our model
using a finite universal set of gates. We show in Section
\ref{qts:sec:probsim} how to efficiently simulate this special case on
unbounded-error probabilistic algorithms.


\subsection{Results and Proof Outline}
\label{qts:sec:result}

Using the notation introduced in Section \ref{qts:sec:classes}, we can
formalize Theorem \ref{qts:thm:maintheorem} and Corollary
\ref{qts:cor:maincorollary} as follows.
\setcounter{backup1}{\value{theorem}}
\setcounter{backup2}{\value{corollary}}
\setcounter{theorem}{1}
\setcounter{corollary}{1}
\addtocounter{theorem}{-1}
\addtocounter{corollary}{-1}
\begin{theorem}[restated]
For every real $d$ and every positive real $\epsilon$ there exists a real
$c>1$ such that either:
\begin{itemize}
\item $\MAJMAJSAT\not\in\BQTIME(n^c)$, or
\item $\MAJSAT\not\in\BQTISP(n^d,n^{1-\epsilon})$.
\end{itemize}
\end{theorem}
\begin{corollary}[restated]
For all $\epsilon>0$, $\MAJMAJSAT\not\in\BQTISP(n^{1+o(1)},n^{1-\epsilon})$.
\end{corollary}
\setcounter{theorem}{\value{backup1}}
\setcounter{corollary}{\value{backup2}}

Theorem \ref{qts:thm:maintheorem} follows immediately from the following two
results. The first gives a lower bound for MajSAT and MajMajSAT on
unbounded-error probabilistic algorithms, and the second translates this lower
bound to the quantum setting by giving a time- and space-efficient simulation
of quantum algorithms by unbounded-error probabilistic algorithms. Recall that
$\PTISP(t,s)$ denotes the class of languages decidable by unbounded-error
probabilistic algorithms running in time $O(t)$ and space $O(s)$.

\begin{lemma}[Allender et al.\ \cite{AKRRV}]
\label{qts:lem:ptisplowerbound}
For every real $d$ and every positive real $\epsilon$ there exists a real $c>1$
such that either:
\begin{itemize}
\item $\MAJMAJSAT\not\in\PTIME(n^c)$, and
\item $\MAJSAT\not\in\PTISP(n^d,n^{1-\epsilon})$.
\end{itemize}
\end{lemma}

\begin{lemma}
\label{qts:lem:main}
For all sufficiently constructible $t\geq\log{n}$ and $s\geq\log{n}$,
\[\BQTISP(t,s)\subseteq\PTISP(t\cdot\polylog(t),\ s+\polylog(t)).\]
\end{lemma}

Lemma \ref{qts:lem:main} is our main technical contribution, and its proof
occupies the remainder of Section \ref{qts:sec:lowerbound}. The first step is
to show that we can assume without loss of generality that our model only uses
a certain finite universal set of quantum gates. A key ingredient is the
Solovay-Kitaev theorem \cite{Kit}, which shows how to approximate any
single-qubit unitary gate to within $\epsilon$ in the $2$-norm sense using
only $\polylog(1/\epsilon)$ gates from a finite universal set. The efficiency
afforded by the Solovay-Kitaev theorem is critical for obtaining our lower
bound.

The second step is to simulate this special case of our model time- and
space-efficiently with unbounded-error probabilistic algorithms. Our strategy
builds on known simulations of quantum computations without intermediate
measurements by probabilistic machines with unbounded error \cite{ADH, FR}.
The basic idea of these simulations is to write the final amplitude of a basis
state as a simple linear combination of $\SHARPP$ functions, where each
$\SHARPP$ function counts the number of quantum computation paths leading to
that state with a certain path amplitude. Taking advantage of our choice of
universal set, we can use simple algebraic manipulations to express the
probability of acceptance as the difference between two $\SHARPP$ functions,
up to a simple common scaling factor. Standard techniques then result in a
time- and space-efficient simulation by an unbounded-error probabilistic
machine.

The above approach only handles unitary operations with one final measurement.
To handle intermediate measurements, we first adapt this approach to capture
the probability of observing any particular sequence of measurement outcomes.
The acceptance probability can then be expressed as a sum over all sequences
of measurement outcomes that lead to acceptance, where each term is the scaled
difference of two $\SHARPP$ functions. We can combine those terms into a
single one using the closure of $\SHARPP$ under uniform exponential sums.
However, the usual way of doing this -- nondeterministically guess and store a
sequence and then run the computation corresponding to that sequence -- is too
space-inefficient. To address this problem, we note that the crux of the
construction corresponds to multiplying two $\SHARPP$ functions on the same
input. The standard approach runs the two computations in sequence, accepting
iff both accept. We argue that we can run these two computations \emph{in
parallel} and keep them in synch so that they access each bit of the guessed
sequence at the same time, allowing us to reference each bit only once. We can
then guess each bit when needed during the final simulation and overwrite it
with the next guess bit, allowing us to meet the space constraint.

Regarding the conditions in Lemma \ref{qts:lem:main}, we assume $t$ and $s$
are at least logarithmic so that they dominate any logarithmic terms arising
from indexed access to the input. We henceforth ignore the technical
constructibility constraints on $t$ and $s$. For Theorem
\ref{qts:thm:maintheorem}, we only need to consider ``ordinary''
polynomially-bounded functions, which are computable in time polynomial in the
length of the output written in binary, which is sufficient for our purposes.


\subsection{Efficient Approximation With a Universal Set}
\label{qts:sec:approx}

A $\BQTISP(t,s)$ computation can be viewed as applying a sequence of $O(t)$
classically selected quantum gates to a register of $O(s)$ qubits. There are
three types of gates:
\begin{itemize}
\item Unitary gates selected from a finite library of gates associated with the
machine.
\item Query gates, which effect the transformation $|i\rangle|b\rangle\mapsto
|i\rangle|b\oplus x_i\rangle$, where $i$ is an index into the input $x$.
\item Measurement gates, which perform a single-qubit projective measurement in
the computational basis.
\end{itemize}

The first step in the proof of Lemma \ref{qts:lem:main} is to show that we
can restrict our attention to machines whose library is a fixed universal
set. It is well-known that every unitary transformation can be effected
exactly using CNOT gates and single-qubit gates. Defining $\BQTISP'(t,s)$ to
be $\BQTISP(t,s)$ with the restriction that each gate in the library either is
CNOT or acts on only one qubit, we have the following.

\begin{lemma}
\label{qts:lem:cnotsingle} For all $t$ and all $s$, \[\BQTISP(t,s)=
\BQTISP'(t,s).\]
\end{lemma}
For completeness, we sketch a proof of Lemma \ref{qts:lem:cnotsingle} in
Appendix \ref{qts:sec:decompose}.

It is also well-known that finite universal sets exist which can approximate
any unitary operation to arbitrary accuracy. We say that a single-qubit
unitary operation $\widetilde{U}$ \emph{$\epsilon$-approximates} a
single-qubit unitary operation $U$ if $||\widetilde{U}-e^{i\theta}U||\leq
\epsilon$ for some (irrelevant) global phase factor $e^{i\theta}$. We say that
a set $S$ of single-qubit unitary gates is \emph{universal for single-qubit
unitary gates} if for all single-qubit unitary gates $U$ and all $\epsilon>0$
there is a sequence $\widetilde{U}_1,\ldots,\widetilde{U}_\ell$ of gates from
$S$ such that the operation $\widetilde{U}_1\cdots\widetilde{U}_\ell$
$\epsilon$-approximates $U$.

We use the fact that the set $\{F,H\}$ is universal for single-qubit unitary
gates\footnote{This is shown on page 196 in \cite{NC} with a relative phase
shift by $\pi/4$ instead of the $F$ gate; however, $F$ is a relative phase
shift by an irrational multiple of $2\pi$ and hence can approximate the
$\pi/4$ phase shift to arbitrary accuracy.}, where $H$ is the Hadamard gate
and \[F=\left[\begin{matrix}1&0\\0&\frac{3}{5}+\frac{4}{5}i\end{matrix}
\right]\] in the computational basis. We can restrict our attention to
quantum Turing machines with library $\{\text{CNOT},F,H\}$ by replacing each
single-qubit gate in the library of a $\BQTISP'$ machine with an approximation
using $F$ and $H$ gates. 

\newpage

We need to satisfy the following requirements:
\begin{itemize}
\item The transformation should not increase the number of gates applied by
too much.
\item The new sequence of gates should still be efficiently computable by a
classical algorithm.
\item The probability an input $x$ is accepted should not change by too much
when we apply the transformation.
\end{itemize}
The following key theorem allows us to meet these constraints.

\begin{lemma}[Solovay and Kitaev \cite{Kit}]
\label{qts:lem:solovaykitaev}
If $S$ is universal for single-qubit unitary gates and is closed under
adjoint, then for all single-qubit unitary gates $U$ and all $\epsilon>0$
there is a sequence of at most $\polylog(1/\epsilon)$ gates from $S$ that
$\epsilon$-approximates $U$. Moreover, such a sequence can be computed
deterministically in time $\polylog(1/\epsilon)$ provided the first $k$ bits
of the matrix entries of $U$ and the gates in $S$ are computable in time
$\poly(k)$.
\end{lemma}

The proof by Solovay and Kitaev gives an algorithm for computing an
approximation in time $\polylog(1/\epsilon)$, ignoring the complexity of
arithmetic (see \cite {DN} and Section 8.3 of \cite{KSV}). We cannot do exact
arithmetic since the entries of our gates may require infinitely many bits to
specify, but in each step of the algorithm it suffices to work with
$\poly(\epsilon)$-approximations to all of the matrices. Computing each matrix
entry to $O(\log{(1/\epsilon)})$ bits suffices for this because the matrices
have only constant size. By our complexity constraint on the transition
amplitudes and the fact that the entries of $F$ and $H$ are also efficiently
computable, this incurs only a $\polylog(1/\epsilon)$ time (and space)
overhead.

We argue that approximating each single-qubit unitary gate to within
$\Theta(1/t)$ ensures that the probability of acceptance of a quantum Turing
machine running in time $t$ only changes by a small amount. Note that our
model is nonoblivious to measurement outcomes, so there may be exponentially
many classical computation paths corresponding to the different measurement
outcomes. A simple union bound over these paths does not work since it would
require exponentially small precision in the approximations, which we cannot
afford. However, the approximation errors are relative to the probability
weights of the paths. As a result, the overall error cannot grow too large.

Define $\BQTISP''(t,s)$ to be $\BQTISP(t,s)$ with the restriction that the
library of gates is $\{\text{CNOT},F,F^\dagger,H,I\}$. We include ``identity
gates'' $I$ for technical reasons -- we need to allow computation steps that
do not change the state of the quantum tape. We include $F^\dagger$ gates
because the Solovay-Kitaev theorem requires the universal set to be closed
under adjoint.

\begin{lemma}
\label{qts:lem:universal}
For all sufficiently constructible $t$ and all $s$, \[\BQTISP'(t,s)\subseteq
\BQTISP''(t\cdot\polylog(t),\ s+\polylog(t)).\]
\end{lemma}
We defer the proof of Lemma \ref{qts:lem:universal} to Appendix
\ref{qts:sec:postpone}. 


\subsection{Efficient Probabilistic Simulation}
\label{qts:sec:probsim}

We now prove Lemma \ref{qts:lem:main}. By Lemmas \ref{qts:lem:cnotsingle} and
\ref{qts:lem:universal}, it suffices to show that \[\BQTISP''(t,s)\subseteq
\PTISP(t,\ s+\log{t}).\] So consider a language $L\in\BQTISP''(t,s)$ and an
associated quantum Turing machine $M$. We fix an arbitrary input $x$ and assume
for simplicity of notation that on input $x$, $M$ uses exactly $s$ qubits
and always applies exactly $t$ quantum gates, exactly $m$ of which are
measurements, regardless of the observed sequence of measurement outcomes.
This can be achieved by padding the computation with gates that do not affect
the accept/reject decision of $M$.


\subsubsection{Computation Tree}

PP can be characterized as the class of languages consisting of inputs for
which the difference of two $\SHARPP$ functions exceeds a certain
polynomial-time computable threshold. Thus, we would like to express the
acceptance probability of $M$ on input $x$ as the ratio of the difference of
two $\SHARPP$ functions and some polynomial-time computable function. To
facilitate the argument, we model the computation of $M$ on input $x$ as a
tree, analogous to the usual computation trees one associates with randomized
or nondeterministic computations. We can express the final amplitude of a
basis state as a linear combination of $\SHARPP$ functions, where each
$\SHARPP$ function counts the number of root-to-leaf paths in the tree that
lead to that basis state and have a particular path amplitude. The
coefficients in this linear combination are the path amplitudes, which are the
products of the transition amplitudes along the path. In order to rewrite the
linear combination as a ratio of the above type, we guarantee certain
properties of the transition amplitudes in the tree.
\begin{itemize}
\item First, our choice of universal set allows us to cancel a common
denominator out of any given gate in such a way that the numerators become
Gaussian integers. We can make the product of the common denominators the
same for all full computation paths, and we will eventually absorb it in the
polynomial-time computable threshold function.
\item Second, the Gaussian integers, such as the $3+4i$ numerator in the $F$
gate, are handled by separating out real and imaginary parts, as well as
positive and negative parts, and by multiplicating nodes such that we
effectively only need to consider numerators in $\{1,-1,i,-i\}$. By
multiplicating nodes we mean that we allow one node to have multiple children
representing the same computational basis state. For example, the $3+4i$
numerator results in seven children.
\end{itemize}

We formally define the computation tree for our fixed input $x$ as follows. It
has depth $t$. Each level $\tau=0,\ldots,t$ represents the state of the
quantum tape after the $\tau$th gate is applied and before the $(\tau+1)$st
gate is applied. Each node $v$ has five labels:
\begin{itemize}
\item $\tau(v)\in\{0,\ldots,t\}$, representing the level of $v$
\item $\mu(v)\in\{0,1\}^{\leq m}$, representing the sequence of measurement
outcomes that leads to $v$
\item $\sigma(v)\in\{0,1\}^s$, representing a computational basis state
\item $\alpha(v)\in\{1,-1,i,-i\}$, representing the numerator of the
amplitude of $v$
\item $\beta(v)\in\mathbb{R}^+$, representing the denominator of the amplitude
of $v$
\end{itemize}
Note that $\alpha(v)$ will be the product of the numerators of the transition
amplitudes along the path that leads to $v$, and similarly for $\beta(v)$.
Labels of nodes across a given level need not be unique; if $v$ and $u$ are at
the same level and $\sigma(v)=\sigma(u)$ and $\mu(v)=\mu(u)$, then $v$ and $u$
represent interference.

We now define the tree inductively as follows. The root node $v$ is at level
$\tau(v)=0$ and has $\mu(v)=\epsilon$ representing that no measurements have
been performed yet, $\sigma(v)=0^s$ representing the initial state, and
$\alpha(v)=\beta(v)=1$ representing that $|0^s\rangle$ has amplitude $1$
initially. Now consider an arbitrary node $v$. If $\tau(v)=t$ then $v$ is a
leaf. Otherwise, $v$ has children at level $\tau(v)+1$ that depend on the type
and operands of $(\tau(v)+1)$st gate applied given that $\mu(v)$ is observed.
Let $G$ denote this gate.

\begin{itemize}
\item If $G=H$ then $v$ has two children $v_0$ and $v_1$. Suppose $G$ is
applied to the $j$th qubit and let $\sigma(v_0)$ and $\sigma(v_1)$ be obtained
from $\sigma(v)$ by setting the $j$th bit to $0$ for $\sigma(v_0)$ and to $1$
for $\sigma(v_1)$. Let $\sigma(v)_j$ denote the $j$th bit of $\sigma(v)$.
If $\sigma(v)_j=0$ then put $\alpha(v_0)=\alpha(v_1)=\alpha(v)$, and if
$\sigma(v)_j=1$ then put $\alpha(v_0)=\alpha(v)$ and $\alpha(v_1)=-\alpha(v)$.
In each case put $\beta(v_0)=\beta(v_1)=\sqrt{2}\beta(v)$. The $1$ and $-1$
multipliers for the $\alpha$-labels correspond to the transition amplitudes of
$G$, except that the common $1/\sqrt{2}$ has been factored out and absorbed in
the $\beta$-label.

\item If $G=F$ then we consider two cases. Suppose $G$ is applied to the $j$th
qubit. If $\sigma(v)_j=1$ then $v$ has seven children, all with the same
$\sigma$-label $\sigma(v)$. Three of the children have $\alpha$-label
$\alpha(v)$ and the other four have $\alpha$-label $i\cdot\alpha(v)$, and all
children have $\beta$-label $5\beta(v)$. This corresponds to an amplitude of
$\frac{3+4i}{5}$, where the numerator $3+4i$ has been spread out across
multiple children so as to maintain the property that all $\alpha$-labels are
in $\{1,-1,i,-i\}$. If $\sigma(v)_j=0$ then $v$ has five children, again all
with the same $\sigma$-label $\sigma(v)$, and now all with the same
$\alpha$-label $\alpha(v)$ and all with $\beta$-label $5\beta(v)$. This
corresponds to an amplitude of $5/5$, so that a common denominator of $5$ can
be used for all nodes resulting from the application of $G$.

\item If $G=F^\dagger$ then the children of $v$ are constructed as in the case
$G=F$ except that the children with $\alpha$-label $i\cdot\alpha(v)$ now have
$\alpha$-label $-i\cdot\alpha(v)$.

\item If $G=I$, CNOT, a query gate, or a measurement gate, then applying $G$
to $|\sigma(v)\rangle$ yields another computational basis state
$|\sigma\rangle$, so $v$ has a single child $u$ with $\sigma(u)=\sigma$,
$\alpha(u)=\alpha(v)$, and $\beta(u)=\beta(v)$.
\end{itemize}
For the cases where $G$ is unitary, we put $\mu(u)=\mu(v)$ for all children
$u$. If $G$ is a measurement gate then we put $\mu(u)=\mu(v)\sigma(v)_j$,
where $u$ is the unique child of $v$ and $j$ is the index of the qubit
measured by $G$.

Note that the denominator $\beta(v)$ can be written as
$5^{f(v)}\sqrt{2}^{h(v)}$, where $f(v)$ denotes the number of $F$ and
$F^\dagger$ gates along the path from the root to $v$, and $h(v)$ the number
of $H$ gates. In fact, $f(v)$ and $h(v)$ can be viewed as functions of
$\tau(v)$ and $\mu(v)$ only, and we will write $f(\tau,\mu)$ and
$h(\tau,\mu)$ accordingly. This is because the sequence of gates that leads
to a node $v$ only depends on $\mu(v)$ (and on the fixed input $x$). The
latter reflects the obliviousness of the model to the quantum computation
path.

In order to describe how the computation tree reflects the evolution
of the quantum tape, we introduce the following notation:
\begin{itemize}
\item $V_{\tau,\mu,\sigma,\alpha}=\{v:\tau(v)=\tau,\ \mu(v)=\mu,\ \sigma(v)=
\sigma,\ \alpha(v)=\alpha\}$
\item $V_{\tau,\mu,\sigma}=\bigcup_\alpha V_{\tau,\mu,\sigma,\alpha}$
\item $V_{\tau,\mu}=\bigcup_\sigma V_{\tau,\mu,\sigma}$
\end{itemize}
Suppose we run $M$ but do not renormalize state vectors after measurements.
Then after $\tau$ gates have been applied, we have a vector for each sequence
of measurement outcomes $\mu$ that could have occurred during the first
$\tau$ steps.
The nodes in $V_{\tau,\mu}$ together with their amplitudes \[\frac{\alpha(v)}
{5^{f(\tau,\mu)}\sqrt{2}^{h(\tau,\mu)}}\] give the vector for $\mu$, since
these are exactly the nodes whose computation paths are consistent with the
measurement outcomes $\mu_1\cdots\mu_{|\mu|}$. More precisely, an inductive
argument shows that the vector for $\mu$ equals \[ \frac{\sum_{v\in V_{\tau,
\mu}}\alpha(v)|\sigma(v)\rangle}{5^{f(\tau,\mu)}\sqrt{2}^{h(\tau,\mu)}}.\]
The squared 2-norm of each such vector equals the probability $p_\mu$
of observing $\mu$. In particular, at the end of the computation we obtain
the following key property.
\begin{claim}
\label{qts:clm:measprob}
For all $\mu\in\{0,1\}^m$, \[p_\mu=\frac{\sum_{\sigma\in\{0,1\}^s}\big|
\sum_{v\in V_{t,\mu,\sigma}}\alpha(v)\big|^2}{25^{f(t,\mu)}2^{h(t,\mu)}}.\]
\end{claim}
We present a formal proof of Claim \ref{qts:clm:measprob} in Appendix
\ref{qts:sec:postpone}.


\subsubsection{Machine Construction}

With Claim \ref{qts:clm:measprob} in hand, we now show how to construct a
probabilistic machine $N$ running in time $O(t)$ and space $O(s+\log{t})$ such
that for all inputs $x$,
\begin{itemize}
\item if $\Pr(\text{$M$ accepts $x$})>1/2$ then
$\Pr(\text{$N$ accepts $x$})>1/2$, and
\item if $\Pr(\text{$M$ accepts $x$})<1/2$ then
$\Pr(\text{$N$ accepts $x$})<1/2$.
\end{itemize}
This suffices to prove Lemma \ref{qts:lem:main}.

We first construct nondeterministic machines $M_1,M_{-1},M_i,M_{-i}$, each
taking as input a triple $(x,\mu,\sigma)$ where $x\in\{0,1\}^n$, $\mu\in
\{0,1\}^m$, and $\sigma\in\{0,1\}^s$. (Recall that most of our notation, such
as $m$ and $s$, is with reference to the particular input $x$.) For each
$\alpha\in\{1,-1,i,-i\}$, $M_\alpha$ will run in time $O(t)$ and space
$O(s+\log{t})$ and satisfy $\#M_\alpha(x,\mu,\sigma)=|V_{t,\mu,\sigma,
\alpha}|$, where $\#M_\alpha(x,\mu,\sigma)$ denotes the number of accepting
computation paths of $M_\alpha$ on input $(x,\mu,\sigma)$. Since $t$ is
constructible, we can assume without loss of generality that all machines are
constructed so as to have exactly $2^g$ computation paths for some
constructible function $g=O(t)$. This allows us to compare numbers of
accepting paths to numbers of rejecting paths.

We simply have $M_\alpha(x,\mu,\sigma)$ nondeterministically guess a
root-to-leaf path in the computation tree. The only information about the
current node $v$ it needs to keep track of is $\sigma(v)$ and $\alpha(v)$,
taking space $O(s)$. It keeps a pointer into $\mu$, taking space $O(\log{t})$.
It determines the correct sequence of gates by simulating the classical part of
$M$, taking $O(t)$ time and $O(s)$ space. When processing a measurement gate
$G$, $M_\alpha$ checks that applying $G$ to the current $\sigma(v)$ yields the
next bit of $\mu$. It rejects if not and otherwise continues, using that bit of
$\mu$ as the measurement outcome. When it reaches a leaf $v$, $M_\alpha$ checks
that $\sigma(v)=\sigma$ and $\alpha(v)=\alpha$ and accepts if so and rejects
otherwise. As constructed, $M_\alpha$ has the desired behavior.

Fix $\mu\in\{0,1\}^m$. By Claim \ref{qts:clm:measprob}, the probability of
observing $\mu$ satisfies
\begin{align*}
p_\mu&=\sum_{\sigma\in\{0,1\}^s}\frac{\big|\sum_{v\in V_{t,\mu,\sigma}}
\alpha(v)\big|^2}{25^{f(t,\mu)}2^{h(t,\mu)}}\\
&=\sum_{\sigma\in\{0,1\}^s}\frac{\big|\sum_\alpha\alpha\cdot\#M_\alpha(x,\mu,
\sigma)\big|^2}{25^{f(t,\mu)}2^{h(t,\mu)}}\\
&=\sum_{\sigma\in\{0,1\}^s}\frac{\Big(\sum_\alpha\#M_\alpha(x,\mu,\sigma)^2
\Big)-\Big(\sum_\alpha\#M_\alpha(x,\mu,\sigma)\cdot\#M_{-\alpha}(x,\mu,\sigma)
\Big)}{25^{f(t,\mu)}2^{h(t,\mu)}}\\
&=\sum_{\sigma\in\{0,1\}^s}\frac{\#M_+(x,\mu,\sigma)-\#M_-(x,\mu,\sigma)}
{25^{f(t,\mu)}2^{h(t,\mu)}},
\end{align*}
where $M_+$ is a nondeterministic machine that guesses $\alpha\in\{1,-1,i,-i\}$
and then runs two copies of $M_\alpha$, accepting iff both accept, and $M_-$ is
a nondeterministic machine that guesses $\alpha\in\{1,-1,i,-i\}$ and then runs
a copy of $M_\alpha$ and a copy of $M_{-\alpha}$, accepting iff both accept. We
run the copies \emph{in parallel}, keeping them in synch so that they access
each bit of $\mu$ at the same time. Note that since $M_+$ and $M_-$ can reject
after seeing a single disagreement with $\mu$, the two copies being run will
apply the same sequence of gates and thus access each bit of $\mu$ at the same
time. It follows that both $M_+$ and $M_-$ need to reference each bit of $\mu$
only once. As we show shortly, this is critical for preserving the space bound.
Both $M_+$ and $M_-$ run in time $O(t)$ and space $O(s+\log{t})$.

In order to capture the probability of acceptance of $M$, we would like to
sum over all complete sequences of measurement outcomes $\mu$ that cause $M$
to accept. We assume without loss of generality that $f(t,\mu)$ and $h(t,\mu)$
are independent of $\mu$, say $f(t,\mu)=f$ and $h(t,\mu)=h$ for all $\mu$, so
that the scaling factor $1/25^{f(t,\mu)}2^{h(\mu)}$ can be factored out of this
sum. To achieve this, we can modify $M$ so that it counts the number of $F$ and
$F^\dagger$ gates and the number of $H$ gates it applies during the computation
and then applies some dummy gates at the end to bring the counts up to the
fixed values $f$ and $h$.

We construct nondeterministic machines $N_+$ and $N_-$, both running in time
$O(t)$ and space $O(s+\log{t})$, such that \[\#N_+(x)=\mathop{\sum_{\text{
$\mu\in\{0,1\}^m$ causing}}}_{\text{$M$ to accept}}\ \sum_{\sigma\in\{0,1\}^s}
\#M_+(x,\mu,\sigma)\] and \[\#N_-(x)=\mathop{\sum_{\text{$\mu\in\{0,1
\}^m$ causing}}}_{\text{$M$ to accept}}\ \sum_{\sigma\in\{0,1\}^s}\#M_-(x,\mu,
\sigma).\] We have $N_+(x)$ run $M_+(x,\mu,\sigma)$ for a nondeterministically
guessed $\mu\in\{0,1\}^m$ and $\sigma\in\{0,1\}^s$ and accept iff $M_+$ accepts
and $\mu$ causes $M$ to accept, and similarly for $N_-$. Since every accepting
execution of $M_+$ or $M_-$ follows an execution of $M$ with measurement
outcomes $\mu$, we know at the end whether $\mu$ causes $M$ to accept.

However, letting $N_+$ just nondeterministically guess $\mu$ and $\sigma$ and
then run $M_+(x,\mu,\sigma)$ does not work because it takes too much space to
store $\mu$. Since $M_+$ and $M_-$ were constructed in such a way that each bit
of $\mu$ is only referenced once, we can nondeterministically guess each
$\mu_j$ when needed and overwrite the previous $\mu_{j-1}$. The space
usage of $\sigma$ is not an issue, so $\sigma$ can be guessed and stored at any
time. Constructed in this way, $N_+$ and $N_-$ have the desired properties.

It follows that the probability $M$ accepts $x$ is \[\frac{\#N_+(x)-\#N_-(x)}
{25^f2^h}.\] Thus,
\begin{itemize}
\item if $\Pr(\text{$M$ accepts $x$})>1/2$ then
$\#N_+(x)-\#N_-(x)>25^f2^{h-1}$, and
\item if $\Pr(\text{$M$ accepts $x$})<1/2$ then
$\#N_+(x)-\#N_-(x)<25^f2^{h-1}$.
\end{itemize}

We can now use a standard technique to obtain the final $\PTISP$ simulation
$N$. We assume without loss of generality that $h\geq 1$. Recall that we assume
$N_+$ and $N_-$ each always have exactly $2^g$ computation paths for some
constructible function $g=O(t)$. By nondeterministically picking $N_+$ or $N_-$
to run, and flipping the answer if $N_-$ was chosen, we get $\big(\#N_+(x)-
\#N_-(x)\big)+2^g$ accepting computation paths. We can generate $2^{g+1}$
dummy computation paths, exactly $2^g+25^f2^{h-1}$ of which reject, to shift
the critical number of accepting paths to exactly half the total number of
computation paths. To do this very time- and space-efficiently, we take
advantage of our use of a universal set, which gives the number of rejecting
dummy paths a simple form. We have $N$ nondeterministically guess $g+1$ bits;
if the first bit is $0$ it rejects, and otherwise it ignores the next $h-1$
bits, groups the next $6f$ bits into groups of $6$ forming a number in
$\{0,\ldots,31\}$, accepts if any group is at least $25$, and otherwise
rejects iff the remaining guess bits are $0$. Since $t$ is constructible, we
can take $f=O(t)$ and $h=O(t)$ to be constructible functions so that $N$ can
compute $f$ and $h$ without affecting the complexity parameters.

As constructed, $N$ runs in time $O(t)$ and space $O(s+\log{t})$ and accepts
$x$ with probability greater than $1/2$ if $x\in L$ and with probability less
than $1/2$ if $x\not\in L$. This finishes the proof of Lemma
\ref{qts:lem:main}.


\subsection{Remarks}
\label{qts:sec:remarks}

Notice that the proof of Lemma \ref{qts:lem:main} shows that the
unbounded-error version of $\BQTISP''(t,s)$ is contained in
$\PTISP(t,\ s+\log{t})$. Since Theorem \ref{qts:thm:maintheorem} only operates
at the granularity of polynomial time and space bounds, it would suffice to
have Lemma \ref{qts:lem:universal} show that $\BQTISP'(t,s)$ is contained in
the unbounded-error version of $\BQTISP''(t^{1+o(1)},\ s+t^{o(1)})$. This
allows us to prove Theorem \ref{qts:thm:maintheorem} under a more relaxed
definition of $\BQTISP(t,s)$:
\begin{itemize}
\item We could relax the error probability to $1/2-1/\poly(t)$ and relax the
time for computing the first $k$ bits of the amplitudes to $2^{o(k)}$, since
the overhead in computing amplitudes would still be subpolynomial and the
Solovay-Kitaev algorithm could still produce $1/\poly(t)$-approximations, for
arbitrarily high degree polynomials, in $t^{o(1)}$ time. The argument in
Lemma \ref{qts:lem:universal} still proves that the error probability remains
less than $1/2$ in this case.

\item Alternatively, we could keep the amplitude efficiency at $\poly(k)$ and
relax the error probability to $1/2-1/2^{t^{o(1)}}$; then the Solovay-Kitaev
algorithm would need to compute $1/2^{t^{o(1)}}$-approximations, which would
still only take $t^{o(1)}$ time.
\end{itemize}

A natural goal is to strengthen Lemma \ref{qts:lem:main} to unbounded-error
quantum algorithms; the problem is that the error probability gets degraded
during the approximation process of Lemma \ref{qts:lem:universal} and thus
needs to be bounded away from $1/2$ by a nonnegligible amount to begin with.

Finally, we remark that nothing prevents our proof of Lemma \ref{qts:lem:main}
from carrying over to any reasonable model of quantum computation that is
nonoblivious to the quantum computation path. In this case, the sequence of
gates leading to a node $v$ in the computation tree does not only depend on
$\mu(v)$, but this fact does not present a problem for our proof. However, the
proof becomes more technical since, e.g., approximating each local unitary
operation may lead to overall nonunitary evolution since the local operations
are themselves applied in a superposition. These complications arise for the
same reason as the unnatural conditions on the transition function in the
models from \cite{BV} and \cite{Wat1}. We feel that working out the details of
such a result would not be well-motivated since the currently envisioned
realizations of quantum computers do not support such nonoblivious behavior.


\section{Conclusion}
\label{qts:sec:concl}

Several questions remain open regarding time-space lower bounds on quantum
models of computation. An obvious goal is to obtain a quantitative improvement
to our lower bound. It would be nice to get a particular constant $c>1$ such
that $\MAJMAJSAT$ cannot be solved by quantum algorithms running in $n^c$ time
and subpolynomial space. The lower bound of Allender et al.\ does yield this;
however, the constant $c$ is very close to $1$, and determining it would
require a complicated analysis involving constant-depth threshold circuitry
for iterated multiplication \cite{Hesse}. Perhaps there is a way to remove the
need for this circuitry in the quantum setting.

A major goal is to prove quantum time-space lower bounds for problems that are
simpler than $\MAJMAJSAT$. Ideally we would like lower bounds for
satisfiability itself, although lower bounds for its cousins in $\PH$ and
$\PARITYP$ would also be very interesting. The difficulty in obtaining such
lower bounds arises from the fact that we know of no simulations of quantum
computations in these classes. The known time-space lower bounds for
satisfiability and related problems follow the indirect diagonalization
paradigm, which involves assuming the lower bound does not hold and then
deriving a contradiction with a direct diagonalization result. For example,
applying this paradigm to quantum algorithms solving $\Sigma_\ell\SAT$ would
entail assuming that $\Sigma_\ell\SAT$ has an efficient quantum algorithm.
Since $\Sigma_\ell\SAT$ is complete for the class $\Sigma_\ell\p$ under very
efficient reductions, this hypothesis gives a general simulation of the latter
class on quantum algorithms. To reach a contradiction with a direct
diagonalization result, we seem to need a way to convert these quantum
computations back into polynomial-time hierarchy computations.

Strengthening Corollary \ref{qts:cor:maincorollary} to $\MAJSAT$ instead of
$\MAJMAJSAT$ may currently be within reach. Recall that the result of
\cite{AKRRV} only needs the following two types of hypotheses to derive a
contradiction:
\begin{itemize}
\item $\MAJMAJSAT\in\PTIME(n^c)$, and
\item $\MAJSAT\in\PTISP(n^d,n^{1-\epsilon})$.
\end{itemize}
Under the hypothesis $\MAJSAT\in\BQTISP(n^{1+o(1)},n^{1-\epsilon})$, Lemma
\ref{qts:lem:main} yields the second inclusion but not the first. One can use
the hypothesis to replace the second majority quantifier of a $\MAJMAJSAT$
formula with a quantum computation. However, we do not know how to use the
hypothesis again to remove the first majority quantifier, because the
hypothesis only applies to majority-quantified \emph{deterministic}
computations. Fortnow and Rogers \cite{FR} prove that $\PP^\BQP=\PP$, and their
proof shows how to absorb the ``quantumness'' into the majority quantifier so
that we \emph{can} apply the hypothesis again. However, their proof critically
uses time-expensive amplification and is not efficient enough to yield a lower
bound for $\MAJSAT$ via the result of \cite{AKRRV}. It might be possible to
exploit the space bound to obtain a more efficient inclusion. It might also be
possible to exploit more special properties of the construction in
\cite{AKRRV} to circumvent the need for the amplification component.


\section*{Acknowledgments}

We would like to thank Scott Diehl for helpful comments.



\appendix

\section{Postponing Measurements}
\label{qts:sec:postpone}

In this appendix we describe a framework for analyzing quantum algorithms
with intermediate measurements by implicitly postponing the measurements
and tracking the unitary evolution of the resulting purification. We stress
that we are doing so for reasons of analysis only; our actual simulations do
not involve postponing measurements. This framework facilitates the proofs of
Claim \ref{qts:clm:measprob} and Lemma \ref{qts:lem:universal}. We first
describe the common framework and then use it for those two proofs.

Consider a quantum Turing machine $M$ running in time $t$ and space $s$. We
fix an arbitrary input $x$ and assume for simplicity of notation that on input
$x$, $M$ uses exactly $s$ qubits and always applies exactly $t$ quantum gates,
exactly $m$ of which are measurements, regardless of the observed sequence of
measurement outcomes. This can be achieved by padding the computation with
gates that do not affect the accept/reject decision of $M$.

We conceptually postpone the measurements in the computation by
\begin{itemize}
\item introducing $m$ ancilla qubits initialized to all $0$'s,
\item replacing the $i$th measurement on each classical computation path by an
operation that entangles the $i$th ancilla qubit with the qubit being measured
(by applying a CNOT to the ancilla with the measured qubit as the control),
and
\item measuring the $m$ ancilla qubits at the end.
\end{itemize}

In the $\tau$th step of the simulation, we apply a unitary operation $U_\tau$
on a system of $s+m$ qubits, where $U_\tau$ acts independently on each of the
subspaces corresponding to distinct sequences of measurement outcomes that can
be observed before time step $\tau$. More precisely, consider the set of
$\mu\in\{0,1\}^{\leq m}$ such that given that $\mu$ is observed, the
$\tau$th gate is applied after $\mu$ is observed but not after the
$(|\mu|+1)$st measurement gate is applied. Let $\mathcal{U}_\tau$ be
the set of $\mu$ such that the $\tau$th gate is unitary, and let
$\mathcal{M}_\tau$ be the set of $\mu$ such that the $\tau$th gate is a
measurement. For $\nu\in\{0,1\}^m$, let $P_\nu$ denote the projection on the
state space of the ancilla qubits to the one-dimensional subspace spanned by
$|\nu\rangle$.

For $\mu\in\mathcal{U}_\tau$, let $G_{\tau,\mu}$ denote the unitary operator
on the state space of $s$ qubits induced by the $\tau$th gate applied given
that $\mu$ is observed. Then $U_\tau$ acts as $G_{\tau,\mu}\otimes I$ on the
range of $I\otimes P_{\mu0^{m-|\mu|}}$. For each $\mu\in\mathcal{M}_\tau$,
$U_\tau$ applies an entangling operation $E_{\tau,\mu}$ that acts only on the
range of $I\otimes (P_{\mu0^{m-|\mu|}}+P_{\mu10^{m-1-|\mu|}})$. The behavior
of $U_\tau$ on the remaining subspaces does not matter; we can set it
arbitrarily to the identity operator. Thus, \[U_\tau=\Big(\sum_{\mu\in
\mathcal{U}_\tau}G_{\tau,\mu}\otimes P_{\mu0^{m-|\mu|}}\Big)+\Big(\sum_{\mu\in
\mathcal{M}_\tau}E_{\tau,\mu}\Big)+R,\] where $R$ is a term that expresses the
behavior on the remaining subspaces.

It is well-known, and can be verified from first principles, that the
probability of observing any sequence of measurement outcomes
$\mu\in\{0,1\}^m$ when $M$ is run equals the probability of observing $\mu$
after the evolution $U=U_tU_{t-1}\cdots U_2U_1$ with all of the ancilla
qubits initialized to $0$. That is, $\Pr(\text{$\mu$ observed})=\big|\big|
(I\otimes P_\mu)U|0^{s+m}\rangle\big|\big|^2$.

We next prove Claim \ref{qts:clm:measprob} and Lemma \ref{qts:lem:universal}.
These proofs both use the above framework but are otherwise independent of
each other.


\subsection{Proof of Claim \ref{qts:clm:measprob}}

Recall that we have a tree expressing the computation of $M$ on the fixed
input $x$, and we wish to show that the probability of observing any complete
sequence of measurement outcomes $\mu\in\{0,1\}^m$ satisfies \[p_\mu=\frac{
\sum_{\sigma\in\{0,1\}^s}\big|\sum_{v\in V_{t,\mu,\sigma}}\alpha(v)\big|^2}
{25^{f(t,\mu)}2^{h(t,\mu)}}.\] Consider the above postponed measurement
framework. The state of the system after $\tau$ steps is given by $U_\tau
\cdots U_1|0^{s+m}\rangle$. We can also write this state as a sum of
contributions from all nodes in the $\tau$th level of the tree. More
precisely, we claim that
\begin{equation}
\label{qts:eqn:decomp}
U_\tau\cdots U_1|0^{s+m}\rangle=\sum_{v\in V_\tau}|\psi(v)\rangle,
\end{equation}
where $V_\tau=\bigcup_\mu V_{\tau,\mu}$ and for each node $v$, \[|\psi(v)
\rangle=\frac{\alpha(v)}{\beta(v)}|\sigma(v)\mu(v)0^{m-|\mu(v)|}\rangle=
\frac{\alpha(v)|\sigma(v)\mu(v)0^{m-|\mu(v)|}\rangle}{5^{f(\tau(v),\mu(v))}
\sqrt{2}^{h(\tau(v),\mu(v))}}.\] Note that $|\psi(v)\rangle$ is the basis
state of $v$ multiplied by its amplitude, with the ancilla qubits set to
indicate the sequence of measurement outcomes that leads to $v$. 

We argue that the decomposition (\ref{qts:eqn:decomp}) holds by induction on
$\tau=0,\ldots,t$. The base case $\tau=0$ is trivial. For $\tau>0$, by
induction it suffices to show that for each node $v\in V_{\tau-1}$,
$U_\tau|\psi(v)\rangle=\sum_{u\in c(v)}| \psi(u)\rangle$, where $c(v)$ denotes
the set of children of $v$. There are two cases. If $\mu(v)\in
\mathcal{U}_\tau$ then it can be verified directly from the construction of
the tree that \[U_\tau|\psi(v)\rangle=\big(G_{\tau,\mu(v)}\otimes
P_{\mu(v)0^{m-|\mu(v)|}}\big)|\psi(v)\rangle=\sum_{u\in c(v)}|\psi(u)
\rangle.\] If $\mu(v)\in\mathcal{M}_\tau$ then it can be directly verified
that $U_\tau|\psi(v)\rangle=E_{\tau,\mu(v)}|\psi(v)\rangle=|\psi(u)\rangle$,
where $u$ is the child of $v$. This completes the induction step.

Given decomposition (\ref{qts:eqn:decomp}) for $\tau=t$, we obtain that for
all $\mu\in\{0,1\}^m$, \[p_\mu=\big|\big|(I\otimes P_\mu)U|0^{s+m}\rangle
\big|\big|^2=\Big|\Big|\sum_{v\in V_{t,\mu}}|\psi(v)\rangle\Big|\Big|^2
=\frac{\sum_{\sigma\in\{0,1\}^s}\big|\sum_{v\in V_{t,\mu,\sigma}}\alpha(v)
\big|^2}{25^{f(t,\mu)}2^{h(t,\mu)}}.\]


\subsection{Proof of Lemma \ref{qts:lem:universal}}

Consider a language $L\in\BQTISP'(t,s)$ and an associated quantum Turing
machine $M'$, and fix an arbitrary input $x$. We assume as above that on
input $x$, $M'$ uses exactly $s$ qubits and always applies exactly $t$ gates,
exactly $m$ of which are measurements. We transform $M'$ into a machine $M''$
running in time $t\cdot\polylog(t)$ and space $s+\polylog(t)$ accepting $L$
with error probability bounded away from $1/2$ by a constant. By standard
amplification techniques, the error probability can be made at most $1/3$, so
$L\in\BQTISP''(t\cdot\polylog(t),\ s+\polylog(t))$.

Using Lemma \ref{qts:lem:solovaykitaev}, we have $M''$ run $M'$ but replace
each single-qubit unitary gate with a $1/20t$-approximation consisting of at
most $\polylog(t)$ gates from the set $\{F,F^\dagger,H\}$. The time and space
overhead is $\polylog(t)$, so $M''$ runs in time $t\cdot\polylog(t)$ and space
$s+\polylog(t)$ and still operates on $s$ qubits. We now show that the
probability $M''$ accepts $x$ differs from the probability $M'$ accepts $x$ by
at most $1/10$. This suffices to prove the lemma.

Let $U'=U'_t\cdots U'_1$ be the evolution on the state space of $s+m$ qubits
obtained by implicitly postponing measurements in the computation of $M'$ as
described above, and let the notation $\mathcal{U}_\tau$ and $G_{\tau,\mu}$ be
as above for this computation. Since the value of $\mu$ uniquely determines
whether $M'$ accepts, we have that $\Pr(\text{$M'$ accepts})=\big|\big|PU'
|0^{s+m}\rangle\big|\big|^2$, where $P$ denotes sum of $I\otimes P_\mu$ over
all $\mu$ consistent with acceptance.

Now let $U''=U''_t\cdots U''_1$ be the same evolution as $U'$ but where each
unitary operation $G_{\tau,\mu}$ is replaced by the operation
$\widetilde{G}_{\tau,\mu}$ that uses the $1/20t$-approximation of the $\tau$th
gate $M'$ applies given that $\mu$ is observed, found by the Solovay-Kitaev
algorithm. Since multiplying by global phase factors does not affect a
computation, we can assume that the approximation used in
$\widetilde{G}_{\tau,\mu}$ is at distance at most $1/20t$ from the original
gate of $M'$. Tensoring with the identity does not change the 2-norm of an
operator, so we also have $\big|\big|\widetilde{G}_{\tau,\mu}-G_{\tau,\mu}
\big|\big|\leq 1/20t$. Now since $U''$ is equivalent to the postponed
measurement transformation applied to $M''$, we have
$\Pr(\text{$M''$ accepts})=\big|\big|PU''|0^{s+m}\rangle\big|\big|^2$.

By standard applications of the triangle inequality (see Box 4.1 in
\cite{NC}), we have that \[\big|\Pr(\text{$M''$ accepts})-\Pr(
\text{$M'$ accepts})\big|\ \leq\ 2\big|\big|U''-U'\big|\big|\ \leq\ 2
\sum_{\tau=1}^t\big|\big|U''_\tau-U'_\tau\big|\big|.\] Thus in order to show
that the acceptance probability of $M'$ and $M''$ on input $x$ differ by at
most $1/10$, it suffices to show that $\big|\big|U''_\tau-U'_\tau\big|\big|
\leq 1/20t$ for all $\tau$. The latter holds since for any unit vector
$|\psi\rangle$ in the state space of $s+m$ qubits, we have
\begin{align*}
\Big|\Big|(U''_\tau-U'_\tau)|\psi\rangle\Big|\Big|^2&=\Big|\Big|\sum_{\mu\in
\mathcal{U}_\tau}\Big(\big(\widetilde{G}_{\tau,\mu}\otimes P_{\mu0^{m-|\mu|}}
\big)-\big(G_{\tau,\mu}\otimes P_{\mu0^{m-|\mu|}}\big)\Big)|\psi\rangle
\Big|\Big|^2\\
&=\sum_{\mu\in\mathcal{U}_\tau}\Big|\Big|\Big(\big(\widetilde{G}_{\tau,\mu}-
G_{\tau,\mu}\big)\otimes I\Big)\big(I\otimes P_{\mu0^{m-|\mu|}}\big)|\psi
\rangle\Big|\Big|^2\\
&\leq\sum_{\mu\in\mathcal{U}_\tau}\Big(\frac{1}{20t}\Big)^2\Big|\Big|\big(
I\otimes P_{\mu0^{m-|\mu|}}\big)|\psi\rangle\Big|\Big|^2\\
&\leq\Big(\frac{1}{20t}\Big)^2.
\end{align*}


\section{Decomposing Quantum Gates}
\label{qts:sec:decompose}

In this appendix we prove Lemma \ref{qts:lem:cnotsingle}, which follows from
results proven in Chapter 4 of \cite{NC}. We include a proof for reasons of
completeness.

For the nontrivial inclusion, consider a language $L\in\BQTISP(t,s)$ and an
associated quantum Turing machine $M$. We convert $M$ into another machine
$M'$ by replacing each application of a library gate $U$ with a sequence of
gates that effects the same operation as $U$, where each either is CNOT or
acts on only one qubit. Then $M'$ accepts an input $x$ with the same
probability as $M$, so to show that $L\in\BQTISP'(t,s)$ we just need to check
that the efficiency parameters are only affected by constant factors and that
the matrix entries of the new gates are still efficiently computable in the
required sense. The time parameter clearly only goes up by a constant factor
that depends on the gates in the library of $M$.

The transformation is done in two steps and uses results proven in Sections
4.3 and 4.5 of \cite{NC}. First, it is shown in $\cite{NC}$ that every unitary
gate can be decomposed as the product of unitary gates each of which acts
nontrivially on only two computational basis states (two-level gates).
Applying this transformation to the library gates associated with the family
does not affect $s$. The matrix entries of these two-level gates are obtained
via standard math operations from the matrix entries of the original gates and
are thus efficiently computable.

Second, it is shown in $\cite{NC}$ that each two-level gate can be decomposed
into a product of CNOTs and single-qubit gates. This transformation can be
done in three steps.
\begin{itemize}
\item First, a two-level gate can be decomposed into a product of operations
each of which is a controlled single-qubit operation that conditions on many
qubits. This is done by using controlled $X$ gates to interchange adjacent
computational basis states in a Gray code order so that the two basis states
acted on by the two-level gate differ only in a single qubit. Then a
controlled single-qubit gate is used to carry out the nontrivial $2\times 2$
submatrix of the two-level gate, and then the basis states are mapped back to
their original values (Figure 4.16 in \cite{NC}). This does not increase the
number of qubits, and all entries in these gates are $0$ or $1$ or come from
the two-level gate, and are hence efficiently computable.

\item Second, each large controlled operation can be reduced to $X$ gates,
Toffoli gates, and a controlled single-qubit gate that conditions on one qubit
(Figure 4.10 in \cite{NC}). To accomplish this, $X$ gates are first used on
some of the control qubits to make the controlled operation condition on all
qubits being $1$. Then these control qubits are ANDed together into some
ancilla qubits using a series of Toffoli gates, and the heart of the operation
is carried out by a controlled single-qubit gate that conditions on the
ancilla qubit holding the AND of the original control qubits. The ANDing
operations are reversed so that the ancilla qubits are reset to $0$ and can
hence be reused and only increase the number of qubits by an additive
constant.

\item Third, the Toffoli and controlled single-qubit gates can be implemented
with special-purpose circuits using only CNOTs and single-qubit gates (Figures
4.9 and 4.6 in $\cite{NC}$). This does not increase the number of qubits, and
the matrix entries in this implementation of a controlled-$U$ gate are
obtained from the matrix entries for $U$ via standard math operations and are
thus efficiently computable.
\end{itemize}

This finishes the proof of Lemma \ref{qts:lem:cnotsingle}.

\end{document}